\documentclass[sigconf, nonacm]{acmart}
\usepackage{placeins}
\usepackage{float}
\AtBeginDocument{%
  }

\setcopyright{acmlicensed}
\copyrightyear{2025}
\acmYear{2025}
\acmConference[KDD Cup Workshop for META CRAG - MM Challenge]{KDD Cup Workshop for META CRAG - MM Challenge}{August 03--07, 2025}{Toronto, Canada}

\begin{document}

\title{Ensembling Multiple Hallucination Detectors Trained \\
on VLLM Internal Representations}

\author{Yuto Nakamizo}
\authornote{Both authors contributed equally to this research.}
\email{nakamizo-y@mm.inf.uec.ac.jp}
\affiliation{%
  \institution{The University of Electro-Communications}
  \state{Tokyo}
  \country{Japan}
}

\author{Ryuhei Miyazato}
\authornotemark[1]
\authornote{Corresponding Author}
\email{miyazato@uec.ac.jp}
\affiliation{%
  \institution{The University of Electro-Communications}
  \state{Tokyo}
  \country{Japan}
}

\author{Hikaru Tanabe}
\email{tanabe-h@mm.inf.uec.ac.jp}
\affiliation{%
  \institution{The University of Electro-Communications}
  \state{Tokyo}
  \country{Japan}
}

\author{Ryuta Yamakura}
\email{yamakura-r@mm.inf.uec.ac.jp}
\affiliation{%
  \institution{The University of Electro-Communications}
  \state{Tokyo}
  \country{Japan}
}

\author{Kiori Hatanaka}
\email{k.hatanaka@uec.ac.jp}
\affiliation{%
  \institution{The University of Electro-Communications}
  \state{Tokyo}
  \country{Japan}
}

\renewcommand{\shortauthors}{Y.Nakamizo, R.Miyazato, H.Tanabe, R.Yamakura, K.Hatanaka}

\begin{abstract}
This paper presents the 5th place solution by our team, y3h2, for the Meta CRAG-MM Challenge at KDD Cup 2025. The CRAG-MM benchmark is a visual question answering (VQA) dataset focused on factual questions about images, including egocentric images. The competition was contested based on VQA accuracy, as judged by an LLM-based automatic evaluator. Since incorrect answers result in negative scores, our strategy focused on reducing hallucinations from the internal representations of the VLLM. Specifically, we trained logistic regression-based hallucination detection models using both the hidden\_state and the outputs of specific attention heads. We then employed an ensemble of these models. As a result, while our method sacrificed some correct answers, it significantly reduced hallucinations and allowed us to place among the top entries on the final leaderboard.
\end{abstract}




\maketitle

\section{Introduction}
Vision-Language Large Models (VLLMs) have made remarkable progress in recent years, significantly enhancing multi-modal understanding and visual question answering (VQA) applications, such as those deployed on smart glasses. However, despite these advances, VLLMs continue to struggle with the generation of hallucinated answers. 
In KDD Cup 2025, Meta introduced the CRAG-MM\cite{crag-mm-2025} benchmark—a VQA dataset focused on factual questions about images, including egocentric images captured with Ray-Ban Meta smart glasses. Participants were tasked with building an MM-RAG system that generates appropriate answers by effectively leveraging external sources and suppressing hallucination.
This paper describes the 5th place solution for the Meta CRAG-MM Challenge at KDD Cup 2025. Our code is available on GitLab\footnote{\url{https://gitlab.aicrowd.com/htanabe/meta-comprehensive-rag-benchmark-starter-kit}}. 



\begin{figure*}[h!]
  \centering
  \includegraphics[width=0.8\linewidth]{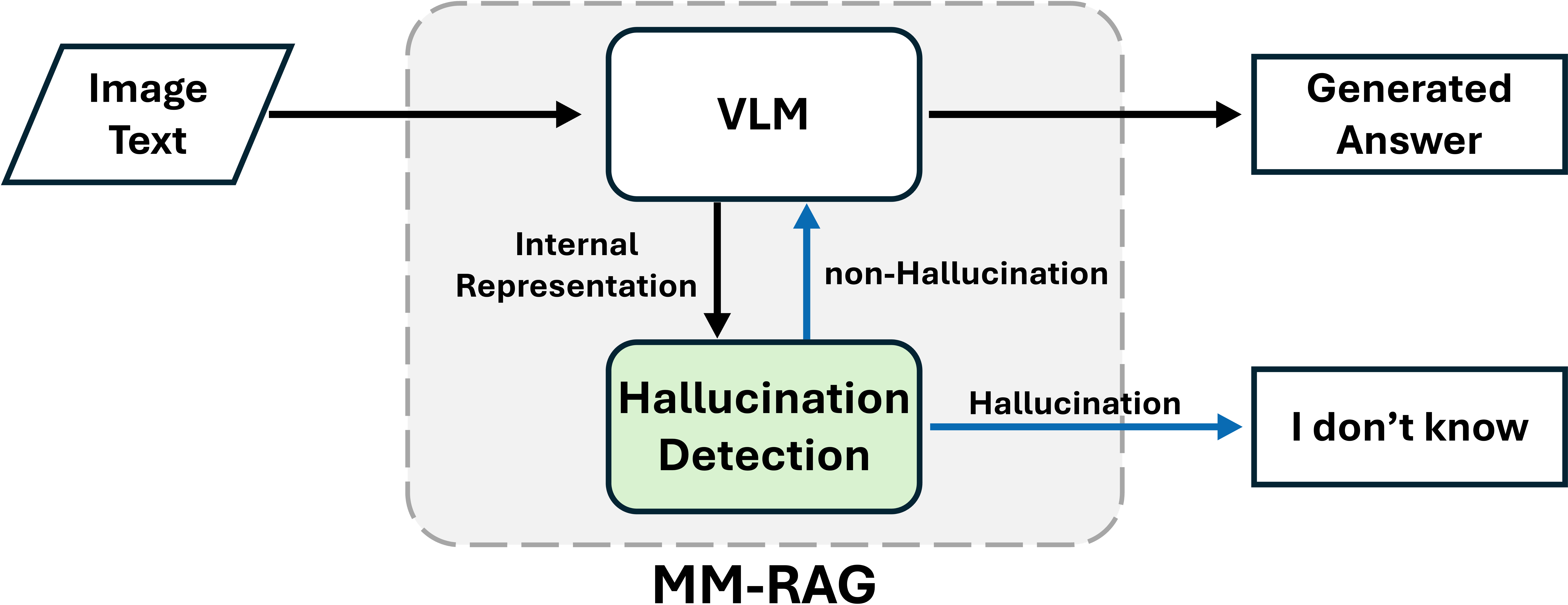}
  \caption{The overview of Our solution}
  \label{fig:overview}
\end{figure*}

\subsection{Dataset}
CRAG-MM is constituted of images paired with questions about the visual content and their corresponding answers. This dataset contains two splits : \emph{validation} and \emph{public\_test}, and they are also labeled this way on Hugging Face. In our paper, we followed this naming convention exactly as given by the organizers.
\begin{itemize}
  \item \textbf{Images: } The dataset includes both egocentric images captured with Meta Ray-Ban glasses and third-person images taken by other individuals.
  \item \textbf{Domain Coverage:} 13 distinct domains (e.g., Books, Food, Science, Shopping, Animals, Vehicles) reflecting diverse real-world scenarios.  
  \item \textbf{QA Category: }
  \begin{itemize}
    \item Simple Recognition (answerable solely from the image)
    \item Simple Knowledge (requires external factual lookup)
    \item Multi-hop questions (requires chaining multiple facts)
    \item Comparison \& Aggregation (requires aggregating or comparing multiple pieces of information)
    \item Reasoning (cannot be directly looked up and require reasoning to answer)  
  \end{itemize}
  \item \textbf{Interaction Modes:} Includes both \emph{single-turn} and \emph{multi-turn} dialogues to evaluate context-aware capabilities.
  \item \textbf{External Sources:}  For all tasks, image-KG-based retrieval is available via an image mock API. Additionally, for Tasks 2 and 3, web retrieval can be performed using a web search mock API.
\end{itemize}

\subsection{Tasks}
The main goal of this competition is to build a Multi-Modal RAG (MM-RAG) system for VQA that utilizes external knowledge sources to achieve higher accuracy on the CRAG-MM benchmark. Specifically, the competition consists of the following three tasks that differ in the types of external knowledge that can be referenced and the forms of interaction.
\begin{enumerate}
  \item \textbf{Task 1: Single-source Augmentation}  \\
  This task evaluates the fundamental answer generation capability of MM-RAG systems. Only the image mock API is provided, which allows retrieval of information associated with similar images.
  
  \item \textbf{Task 2: Multi-source Augmentation}  \\
  This task evaluates the system's ability to synthesize information from multiple sources by additionally providing a mock web search API. Participants are required to integrate information from both the image-KG and web sources to generate answers.
  
  \item \textbf{Task 3: Multi-turn QA}  \\
  This task assesses the system’s ability to handle multi-turn conversations, requiring not only accurate question answering but also contextual understanding and conversation smoothness across multiple turns.
\end{enumerate}

\subsection{Evaluation Metrics}
\begin{itemize}
    \item \textbf{Automatic Grading:}
    \begin{itemize}
        \item Answers are graded by an LLM-based automatic judge for leaderboard.
        \item Each answer is classified as \emph{correct}, \emph{partially correct}, \emph{incorrect}, or \emph{missing/refusal}.
    \end{itemize}
    \item \textbf{Manuel Grading:}
    \begin{itemize}
        \item Answers that receive high scores on automatic judge are subsequently re‑evaluated by human assessors to determine the final ranking.
        \item Each answer is classified as \emph{correct}, \emph{partially correct}, \emph{incorrect}, or \emph{missing/refusal}.
    \end{itemize}
    \item \textbf{Scoring:}
    \begin{itemize}
        \item $+1$ point for a correct answer.
        \item $0$ points for partial, missing, or refusal.
        \item $-1$ point for an incorrect (hallucinated or wrong) answer.
        \item Leaderboard score is determined by trustfulness score, which is the average score across all examples in the evaluation set for a given MM-RAG system.
    \end{itemize}
\end{itemize}

\subsection{models}
It is required to use Llama models to build MM-RAG in this competition. Especially, participants can use or fine-tune the Llama 3 models, which can be run on 48GB GPU memory. In our MM-RAG, we adapted \textbf{meta-llama/Llama-3.2-11B-Vision-Instruct} \footnote{\url{https://huggingface.co/meta-llama/Llama-3.2-11B-Vision-Instruct}}

\section{Solution}

In this competition, the performance of MM-RAG is evaluated by the average score, where a correct answer receives +1, an incorrect answer (hallucination) receives -1, and "I don't know" yields 0. The final score, called the trustfulness score, is the average score across all VQA pairs. Since a high number of incorrect answers leads to a negative trustfulness score, we adopted a strategy that detects hallucinations from internal representations of VLLM, which is used to filter answers: whenever a hallucination is detected, the system returns "I don't know." Previous studies (e.g., \cite{li2023inferencetime, du2024haloscope, yang2025mitigating}) have shown that hallucination generation in LLMs or VLLMs can be detected from internal representations, and in particular, the outputs of attention heads are known to be effective for hallucination detection. We applied these research findings into our solution. \par
Figure~\ref{fig:overview} presents an overview of our solution. In our solution, we simply applied a baseline model (LlamaVisionModel) equipped with a hallucination detection filter to all tasks. \par
The other approaches we tried for mitigating hallucination are described in the Appendix~\ref{appendix:other_approaches}.

\label{sec:hallucination_detection}
\begin{figure*}[h!]
  \centering
  \includegraphics[width=0.9\linewidth]{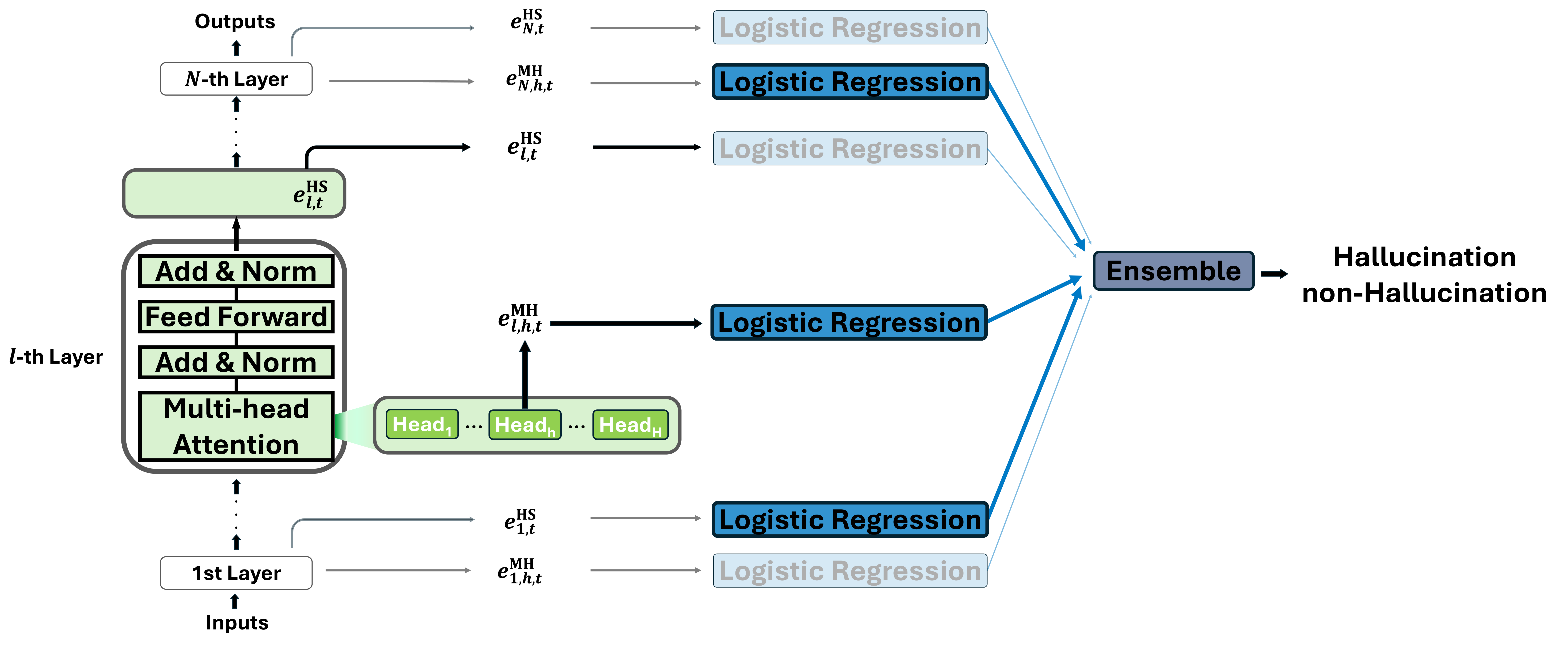}
  \caption{Hallucination Detection Filter}
  \label{fig:hallucination_detection}
\end{figure*}

\subsection{Hallucination Detection Filter}

\begin{figure}[h!]
  \centering
  \includegraphics[width=0.8\linewidth]{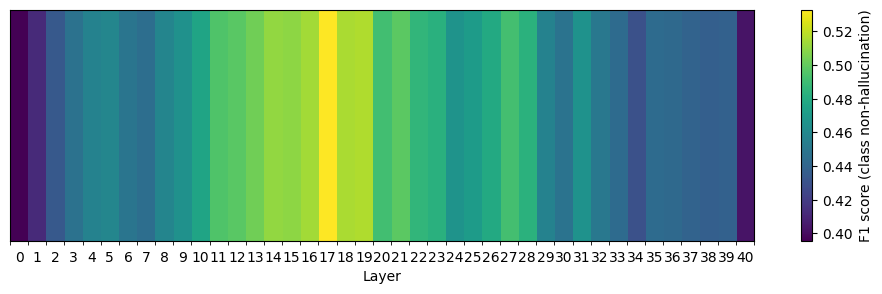}
  \caption{F1 score for hallucination detection from hidden state}
  \label{fig:f1_hs}
\end{figure}

\begin{figure}[h!]
  \centering
  \includegraphics[width=0.8\linewidth]{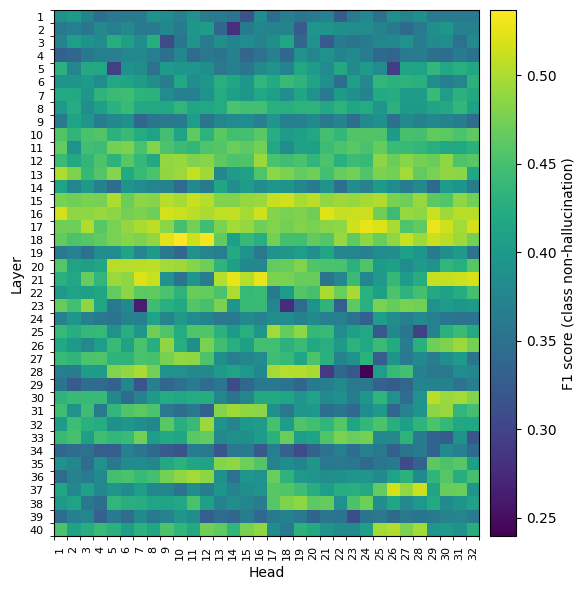}
  \caption{F1 score for hallucination detection from attention heads}
  \label{fig:f1_ah}
\end{figure}

Our goal is to detect whether the answer being generated by the VLLM during inference is correct or a hallucination, based on its internal representations, and to automatically return "I don't know" whenever a hallucination is detected. \par
Specifically, we trained logistic regression models to classify answers as hallucinations or not, using the hidden states from each layer (hidden\_states) and the outputs from each attention head (attention\_outputs) as features. The overview of Hallucination Detection Model is shown in Figure~\ref{fig:hallucination_detection}. \par
First, we extracted the internal representations by getting model to answer VQA from the \emph{validation} split of the dataset. For each predicted token $t$, we extracted the final token’s activation from both the hidden state ($e^{HS}_{l,t}$) and the attention head ($e^{MH}_{l,h,t}$). We computed the average $e_t$ of generated across all token predictions to obtain $\mathbf{e}$, which is then used as the feature representation for each layer's hidden state and attention head. Concurrently, we evaluated the generated answer using LLM-as-a-Judge, labeling correct and incorrect answers as $y = 1, 0$, respectively. Using these $\mathbf{X}$ and $y$, we trained logistic regression models: 
\begin{equation}
    \widehat{y} = \sigma(\mathbf{w}^\top \mathbf{e} + b)
\end{equation}
where $\sigma(\cdot)$ is the sigmoid function, $\mathbf{w}$ is the weight vector, $\mathbf{e}$ is the input feature vector (either $e^{HS}$ or $e^{MH}$), and $b$ is the bias term. Here, the dimensionality of $e_{HS}$ was reduced using PCA, retaining as many principal components as needed to ensure that the cumulative contribution rate exceeded 95\%. \par

Next, we used the data of \emph{public\_test} split and select which layer’s hidden state and attention head outputs to use for hallucination detection. While model answering on the \emph{public\_test} set, we applied the trained hallucination detection models. Then, we selected the hidden states and attention heads that could sufficiently detect hallucinations, specifically those with a non-hallucination F1 score above 0.5. This threshold was chosen empirically. The detection performance for each hidden state and attention head is shown in Figure~\ref{fig:f1_hs} and Figure~\ref{fig:f1_ah}. We selected 7 hidden states and 58 attention heads, and trained a total of 65 logistic regression models. The selected layers and positions are described in the Appendix~\ref{appendix:selected_layer}. \par

\begin{table*}[h!]
\centering
\caption{Results for Task 1, Task 2, and Task 3. Each table shows accuracy, missing, hallucination, and trustfulness.}

\begin{tabular}{l|c|c|c|c}
\multicolumn{5}{c}{\textbf{Single-turn tasks}} \\
\hline
Method & Accuracy & Missing & Hallucination & Trustfulness \\
\hline
Baseline & $0.207$ & $0.058$ & $0.735$ & $-0.528$ \\
+ only hidden\_state (HS) & $0.099\ (\downarrow 52\%) $ & $0.812$ & $0.089\ (\downarrow 88\%) $ & $0.010$ \\
+ only attention\_heads (MH) & $0.081\ (\downarrow 61\%)$ & $0.873$ & $0.046\ (\downarrow 94\%) $ & $0.035$ \\
+ HS+ MH (our method)& $0.082\ (\downarrow 61\%) $ & $0.873$ & $0.045\ (\downarrow 94\%) $ & $0.036$ \\ \hline
Leaderboard (Task 1) & $0.088$ & $0.860$ & $0.052$ & $0.036$ \\
Leaderboard (Task 2) & $0.088$ & $0.859$ & $0.053$ & $0.034$ \\
\hline
\end{tabular}

\vspace{1.5em}

\begin{tabular}{l|c|c|c|c}
\multicolumn{5}{c}{\textbf{Multi-turn task}} \\
\hline
Method & Accuracy & Missing & Hallucination & Trustfulness \\
\hline
Baseline & $0.268$ & $0.352$ & $0.379$ & $-0.111$ \\
+ only hidden\_state (HS) & $0.173\ (\downarrow 36\%)$ & $0.742$ & $0.084\ (\downarrow 78\%)$ & $0.089$ \\
+ only attention\_heads (MH) & $0.156\ (\downarrow 42\%)$ & $0.784$ & $0.060\ (\downarrow 84\%)$ & $0.097$ \\
+ HS+ MH (Our Method)& $0.155\ (\downarrow 42\%)$ & $0.784$ & $0.061\ (\downarrow 84\%)$ & $0.094$ \\ \hline
Leaderboard (Task 3) & 0.138 & 0.827 & 0.035 & 0.104 \\
\hline
\end{tabular}
\label{tab:result}
\end{table*}

Finally, we ensembled the detection models trained on the selected hidden states and attention heads by averaging their outputs. 

For each sample, we calculated the mean of the predictions $\widehat{y}j$ from all ensembled models:
\begin{equation}
\overline{y} = \frac{1}{N} \sum_{j=1}^{N} \widehat{y}_j
\end{equation}
where $N$ is the number of ensembled models.

\subsection{Visual-Question Answering}

\begin{figure}[H]
  \centering
  \includegraphics[width=0.8\linewidth]{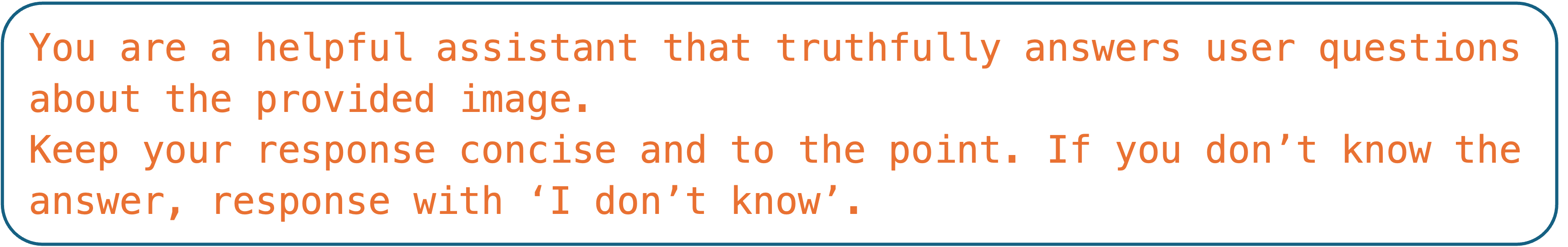}
  \caption{The prompt used in Our Solution}
  \label{fig:prompt}
\end{figure}

An overview of our solution is presented in Figure~\ref{fig:overview}.
We applied the Hallucination Detection Filter trained in Section~\ref{sec:hallucination_detection} to the baseline shared by the organizers, LlamaVisionModel. For the prompt, we directly used the one used in the baseline, shown in Figure~\ref{fig:prompt}. No external sources were used at all. \par

When an image and question are given as input, the VLLM attempts to generate an answer. At this stage, the hallucination detection model trained in Section~\ref{sec:hallucination_detection} is used to judge, based on the internal activations, whether the generated answer is a hallucination. Specifically, if the predicted value for $y=0$ (non-hallucination) exceeds 0.65, the answer is accepted as is; otherwise, the system outputs "I don't know." The threshold value of 0.65 was chosen by selecting the value that yielded the highest trustfulness score on the \emph{public\_test} split.


\section{Results}
The results of our strategy are shown in Table~\ref{tab:result}. In the upper table, we report the performance on the single‐turn tasks (Task 1 and Task 2). The baseline model achieved an accuracy of $0.207$ and a hallucination rate of $0.735$ (Trustfulness = $-0.528$). Applying the hidden‐state filter (+ only hidden\_state) reduced accuracy to $0.099\ (\downarrow 52\%)$ but suppressed hallucinations to $0.089\ (\downarrow 88\%)$ and improved Trustfulness to 0.010. The attention‐head filter (+ only attention\_heads) further lowered accuracy to $0.081\ (\downarrow 61\%)$, but cut hallucinations to $0.046\ (\downarrow 94\%)$, yielding a Trustfulness of $0.035$. Our combined method (HS+AH) maintained a high level of hallucination suppression of $0.045\ (\downarrow 94\%)$ and achieved the highest offline reliability of $0.036$, although the correct response rate also dropped to $0.082$. On the leaderboard’s private test sets, our approach yielded an accuracy of $0.088$, a hallucination rate of $0.052$ for Task 1 and $0.053$ for Task 2, and Trustfulness scores of $0.036$ and $0.034$, respectively.

In the lower table, we report performance on the multi-turn task (Task 3). The baseline model achieved an accuracy of $0.268$, a missing rate of $0.352$, and a hallucination rate of $0.379$ (Trustfulness = $-0.111$). Applying the hidden-state filter (+ only hidden-state) reduced accuracy to $0.173\ (\downarrow 36\%)$ but suppressed hallucinations to $0.084\ (\downarrow 78\%)$ and improved Trustfulness to $0.089$. The attention-head filter (+ only attention\_heads) further lowered accuracy to $0.156\ (\downarrow 42\%)$ and cut hallucinations to $0.060\ (\downarrow 84\%)$, yielding a Trustfulness of $0.097$. Our combined method (HS+AH) maintained this high level of hallucination suppression, with a hallucination rate of $0.061\ (\downarrow 84\%)$ at an accuracy of $0.155\ (\downarrow 42\%)$, achieving a Trustfulness of $0.094$. Supplementary analysis conducted during manuscript preparation revealed that, for the multi-turn task, the attention-head filter alone actually achieved the highest Trustfulness, slightly surpassing the combined method’s, but in the competition, we unified with the single-turn and submitted the combined method. On the leaderboard’s private test set for Task 3, our approach yielded an accuracy of $0.138$, a hallucination rate of $0.035$, and a Trustfulness of $0.104$.

An ablation study on filtering is additionally provided in the Appendix~\ref{appendix:f1_threshold}.

\section{Conclusion}
In the META CRAG-MM Challenge, it was essential not only to improve VQA accuracy but also to address the challenge of hallucinations. While leveraging external sources through RAG can enhance factual correctness, it does not fully eliminate the risk of generating hallucinated answers. Building more trustworthy AI involves not only expanding the range of a model’s knowledge, but also enabling it to recognize and explicitly indicate when it lacks the necessary information to answer a question. Our experience in this KDD Cup reinforced the importance of such mechanisms for developing reliable and responsible AI systems. In the future, we aim to focus on developing LLMs that can reliably recognize and admit when they do not know the answer.

\begin{acks}
We would like to express our sincere gratitude to Meta, AIcrowd, and all the KDD Cup organizers. This competition provided a valuable opportunity to deepen our understanding of VLLMs and MM-RAG.  \par
RM work was supported by JST K Program Japan Grant Number JPMJKP24C3.  \par
\end{acks}

\bibliographystyle{ACM-Reference-Format}
\bibliography{main}

\newpage
\appendix
\onecolumn

\section{Other attempted approaches}
\label{appendix:other_approaches}
To mitigate hallucination, we also tried several other methods.
\subsection{Prompt Control}
\begin{figure}[H]
  \centering
  \includegraphics[width=0.9\linewidth]{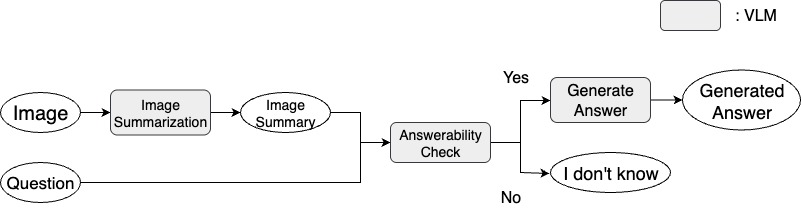}
  \caption{The overview of Answerability Check}
  \label{fig:ac}
\end{figure}
\vspace{-2mm}
\begin{figure}[H]
  \centering
  \includegraphics[width=0.9\linewidth]{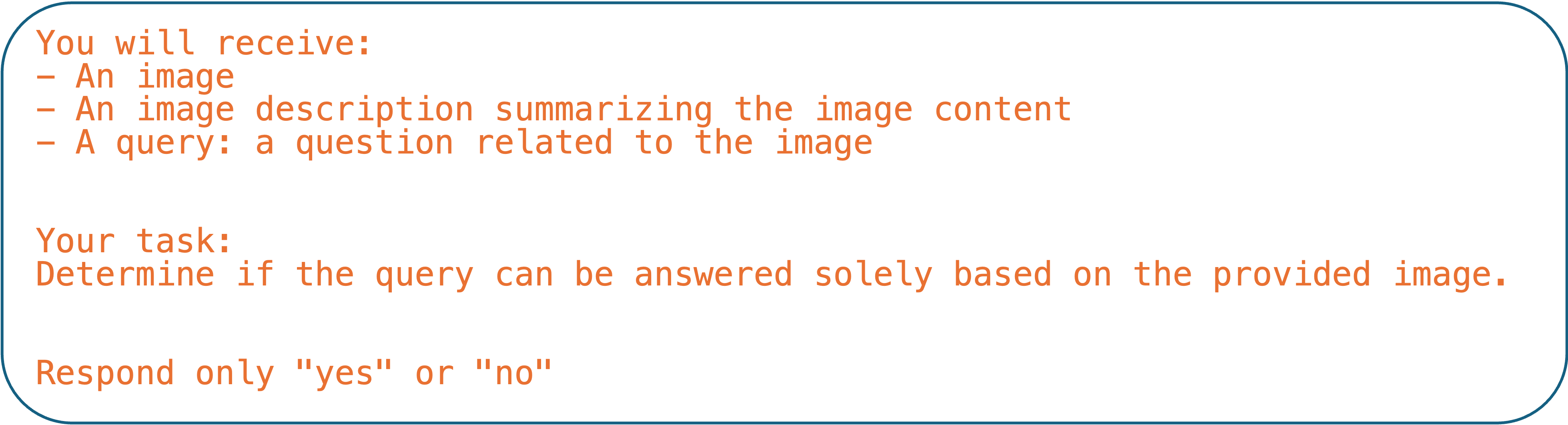}
  \caption{The prompt for image summarization}
  \label{fig:prompt_is}
\end{figure}
\begin{figure}[H]
  \centering
  \includegraphics[width=0.9\linewidth]{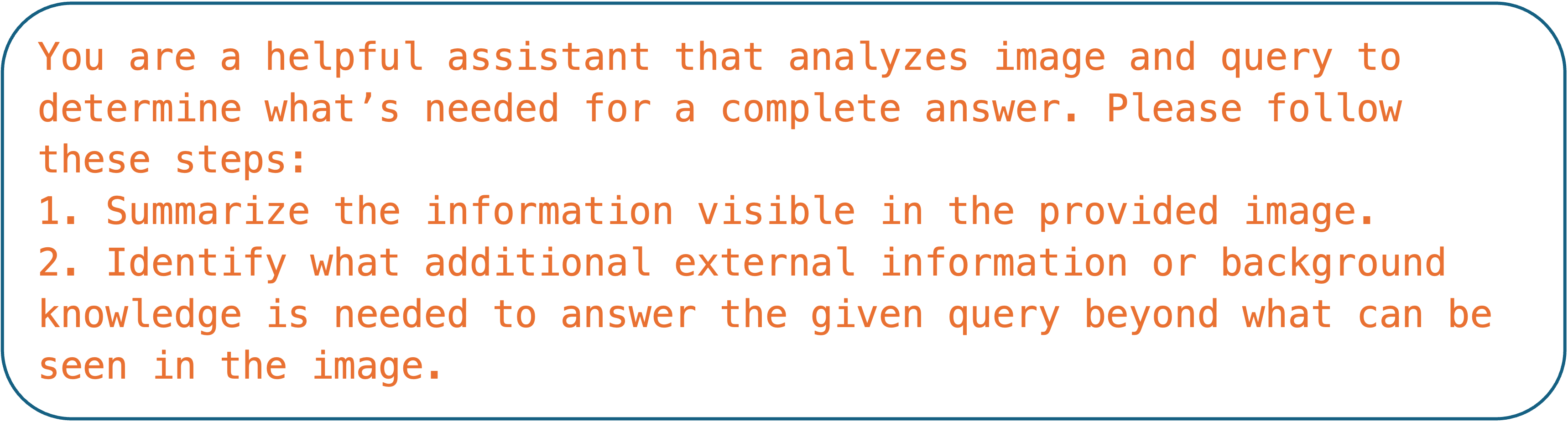}
  \caption{The prompt for answerability check}
  \label{fig:prompt_ac}
\end{figure}
\begin{figure}[H]
  \centering
  \includegraphics[width=0.9\linewidth]{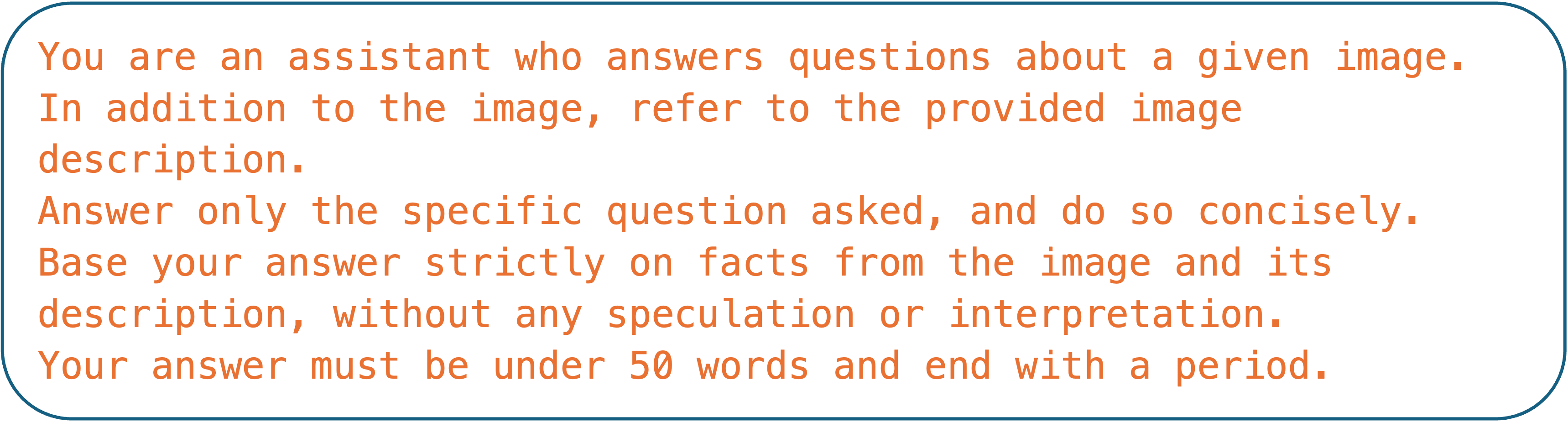}
  \caption{The prompt for generating answer}
  \label{fig:prompt_ga}
\end{figure}
\begin{itemize}
    \item \textbf{Answerability Check (AC)} (Figure~\ref{fig:ac}):  \\
    Given an image and a question, the VLLM produces an image summary (Prompt: Figure~\ref{fig:prompt_is}). An answerability check then determines whether the question can be answered based on the image summary and question (Prompt: Figure~\ref{fig:prompt_ac}). If so, the model generates an answer Prompt: Figure~\ref{fig:prompt_ga}; otherwise, it outputs “I don’t know.”
\end{itemize}


\subsection{Internal states of VLLM}
\begin{itemize}
    \item \textbf{Logprobs} \\
    During answer generation, we extracted the logprobs. When the logprob of the generated answer was below the threshold, the model was forced to output "I don’t know". We report results with thresholds of $-0.01$, $-0.07$, and $-0.1$. 
\end{itemize}

\subsection{Results}
\begin{table}[H]
\centering
\caption{The results of other attempted approach}
\label{tab:results_others}
\begin{tabular}{l|c|c|c|c}
\hline
Method & Accuracy & Missing & Hallucination & Trustfulness \\
\hline
Baseline & $0.207$ & $0.058$ & $0.735$ & $-0.528$ \\
AC & $0.080$ & $0.802$ & $0.118$ & $-0.039$ \\ \hline
Logprobs (-0.01) & $0.063$ & $0.853$ & $0.084$ & $-0.021$\\
Logprobs (-0.07)& $0.064$ & $0.853$ & $0.084$ & $-0.020$ \\
Logprobs (-0.1) & $0.064$ & $0.848$ & $0.089$ & $-0.025$ \\ \hline \hline
Our Method & $0.088$ & $0.860$ & $0.052$ & $0.036$ \\
\hline
\end{tabular}
\end{table}

Although we tried several alternative methods as shown in the Table~\ref{tab:results_others}, the approach of training a hallucination detection model on internal activations (Our Method) worked best, so we adopted it as our final solution.

\section{The selected layers and head positions}
\label{appendix:selected_layer}

\begin{table}[h]
\centering
\caption{The used hidden state selected by non-hallucination F1 score.}
\begin{tabular}{c|c|c}
\hline
Rank & Layer & F1 \\
\hline
1 & 17 & 0.5329 \\
2 & 19 & 0.5167 \\
3 & 18 & 0.5154 \\
4 & 16 & 0.5139 \\
5 & 14 & 0.5109 \\
6 & 15 & 0.5093 \\
7 & 13 & 0.5032 \\
\hline
\end{tabular}
\end{table}
\FloatBarrier
\begin{table}[H]
\centering
\small
\caption{Top 58 attention heads selected by non-hallucination F1 score.}
\begin{tabular}{c|c|c|c||c|c|c|c}
\hline
Rank & Layer & Head & F1 & Rank & Layer & Head & F1 \\
\hline
1 & 17 & 9 & 0.5368 & 30 & 17 & 26 & 0.5118 \\
2 & 17 & 11 & 0.5317 & 31 & 12 & 10 & 0.5109 \\
3 & 16 & 28 & 0.5278 & 32 & 15 & 21 & 0.5109 \\
4 & 20 & 15 & 0.5271 & 33 & 20 & 28 & 0.5108 \\
5 & 16 & 23 & 0.5256 & 34 & 14 & 16 & 0.5107 \\
6 & 20 & 13 & 0.5252 & 35 & 20 & 29 & 0.5102 \\
7 & 17 & 8 & 0.5239 & 36 & 15 & 13 & 0.5093 \\
8 & 15 & 20 & 0.5217 & 37 & 15 & 12 & 0.5078 \\
9 & 20 & 6 & 0.5216 & 38 & 16 & 7 & 0.5069 \\
10 & 16 & 24 & 0.5198 & 39 & 19 & 7 & 0.5066 \\
11 & 20 & 31 & 0.5182 & 40 & 14 & 19 & 0.5063 \\
12 & 36 & 25 & 0.5177 & 41 & 19 & 6 & 0.5062 \\
13 & 16 & 31 & 0.5169 & 42 & 29 & 28 & 0.5060 \\
14 & 15 & 8 & 0.5167 & 43 & 15 & 10 & 0.5058 \\
15 & 15 & 0 & 0.5166 & 44 & 17 & 29 & 0.5054 \\
16 & 15 & 15 & 0.5157 & 45 & 14 & 8 & 0.5051 \\
17 & 14 & 17 & 0.5157 & 46 & 15 & 30 & 0.5050 \\
18 & 16 & 29 & 0.5156 & 47 & 15 & 31 & 0.5047 \\
19 & 16 & 22 & 0.5151 & 48 & 19 & 4 & 0.5047 \\
20 & 17 & 10 & 0.5137 & 49 & 19 & 5 & 0.5041 \\
21 & 20 & 30 & 0.5136 & 50 & 20 & 14 & 0.5040 \\
22 & 15 & 23 & 0.5134 & 51 & 27 & 17 & 0.5040 \\
23 & 36 & 27 & 0.5133 & 52 & 20 & 12 & 0.5040 \\
24 & 17 & 28 & 0.5131 & 53 & 14 & 11 & 0.5035 \\
25 & 15 & 9 & 0.5128 & 54 & 27 & 18 & 0.5021 \\
26 & 15 & 22 & 0.5127 & 55 & 12 & 0 & 0.5018 \\
27 & 14 & 10 & 0.5121 & 56 & 14 & 24 & 0.5013 \\
28 & 15 & 28 & 0.5120 & 57 & 16 & 2 & 0.5009 \\
29 & 20 & 7 & 0.5119 & 58 & 39 & 25 & 0.5009 \\
\hline
\end{tabular}
\end{table}

\section{F1-based Filter Selection Ablation}
\label{appendix:f1_threshold}
\begin{table*}[htpb]
\centering
\caption{Maximum Accuracy/Hallucination/Trustfulness and number of selected filters at each F1 threshold}
\label{tab:results_f1_threshold}
\begin{tabular}{c|c|c|c|c|c|c}
\hline
F1 threshold & Number of filters & Accuracy & Missing & Hallucination & Trustfulness & Ensemble threshold\\
\hline
0.0 & $1321$ & $0.068$ & $0.886$ & $0.045$ & $0.023$ & $0.56$ \\
0.1 & $1321$ & $0.068$ & $0.886$ & $0.045$ & $0.023$ & $0.56$ \\
0.2 & $1321$ & $0.068$ & $0.886$ & $0.045$ & $0.023$ & $0.56$ \\
0.3 & $1313$ & $0.069$ & $0.885$ & $0.046$ & $0.023$ & $0.56$ \\
0.4 & $752$ & $0.074$ & $0.878$ & $0.048$ & $0.026$ & $0.58$ \\
0.5 & $65$ & $0.082$ & $0.873$ & $0.045$ & $0.036$ & $0.65$ \\
\hline
\end{tabular}
\end{table*}
We conducted an additional investigation into the effect of varying the F1-score threshold in filter selection.
For each threshold setting (ranging from 0.0 to 1.0 in increments of 0.1), the selection criterion was adjusted such that only filters with an F1 score strictly exceeding the threshold were retained.
For each case, we report in Table~\ref{tab:results_f1_threshold} the the best evaluation scores, together with the corresponding number of selected filters and the threshold applied for the final hallucination judgment.
When the F1 threshold was set to 0.6 or higher, no filters met the condition.
Therefore, no results are reported for these ranges.
Among the evaluated settings, a threshold of 0.5 yielded the best performance, and this value was adopted in our final competition submission.

\end{document}